\documentclass[useAMS,usenatbib,fleqn]{mn2e}
\usepackage{amsmath, array, booktabs, xifthen, graphicx, color, xspace, url, amssymb, mathtools, enumitem, natbib, physics, listings, topcapt}

\usepackage{tikz,xcolor}

\usepackage[normalem]{ulem}

\usepackage[pagebackref=false]{hyperref}
\hypersetup{
    colorlinks=true,
    citecolor=blue,
    linkcolor=black,
    urlcolor=black,
    pdfpagemode=FullScreen,
}

\def \aap{A\&A}

\def \apjl{ApJ}
\def \apjs{ApJS}

\setlist[enumerate]{wide=0pt, leftmargin=15pt, labelwidth=15pt, align=left}

\newcommand{\reb}{{\sc \tt REBOUND}\xspace}
\newcommand{\rebx}{{\sc \tt REBOUNDx}\xspace}
\newcommand{\whfast}{{\sc \tt WHFast}\xspace}
\newcommand{\whckl}{{\sc \tt WHCKL}\xspace}

\newcommand{\bo}[1][]{%
    \ifthenelse{\equal{#1}{}}{\mathcal{O}}{\mathcal{O}\left(#1\right)}%
}

\NewDocumentCommand{\code}{v}{%
\texttt{{#1}}%
}

\newcommand{\fig}[1]{Fig.~\ref{#1}}
\newcommand{\eq}[1]{equation~(\ref{#1})}

\title[General relativity and long-term stability]{General relativistic precession and the long-term stability of the solar system}
\date{Draft version: \today{}}

\author[Brown \& Rein]{Garett Brown$^{1,2\;\star}$ and 
    Hanno Rein$^{1,2,3}$\\
$^1$ Department of Physical and Environmental Sciences, University of Toronto at Scarborough, Toronto, Ontario M1C 1A4, Canada,\\
$^2$ Department of Physics, University of Toronto, Toronto, Ontario, M5S 3H4, Canada,\\
$^3$ Department of Astronomy and Astrophysics, University of Toronto, Toronto, Ontario, M5S 3H4, Canada\\
$\star\;$ \rm{garett.brown@mail.utoronto.ca}\\
}

\vspace{0.5\baselineskip}

\begin{document}
\maketitle

\begin{abstract}
The long-term evolution of the solar system is chaotic. 
In some cases, chaotic diffusion caused by an overlap of secular resonances can increase the eccentricity of planets when they enter into a linear secular resonance, driving the system to instability. 
Previous work has shown that including general relativistic contributions to the planets' precession frequency is crucial when modelling the solar system. 
It reduces the probability that the solar system destabilizes within 5 Gyr by a factor of 60. 
We run 1280 additional $N$-body simulations of the solar system spanning 12.5 Gyr where we allow the GR precession rate to vary with time. 
We develop a simple, unified, Fokker-Planck advection-diffusion model that can reproduce the instability time of Mercury with, without, and with time-varying GR precession. 
We show that while ignoring GR precession does move Mercury's precession frequency closer to a resonance with Jupiter, this alone does not explain the increased instability rate. 
It is necessary that there is also a significant increase in the rate of diffusion. 
We find that the system responds smoothly to a change in the precession frequency: There is no critical GR precession frequency below which the solar system becomes significantly more unstable. 
Our results show that the long-term evolution of the solar system is well described with an advection-diffusion model. 
\end{abstract}


\section{Introduction}
\label{sec:intro}

For many centuries, the motion of the solar system has been a subject of interest to many cultures.
Since \cite{Newton1687} formulated his universal law of gravitation, there has been a formal and precise way to describe this motion.
The paradigm shift of general relativity (GR) introduced by \cite{Einstein1915} further increased our understanding and ability to model the motion of the solar system precisely.
Shortly afterwards, post-Newtonian corrections derived from the equations of general relativity provided a simpler approach to more accurately model the discrepancies previously found through observation~\citep{Eddington1923}.
Today, a combination of these precise numerical models together with modern computational resources have made extensive studies on the long-term behaviour of the solar system feasible.
Modern numerical studies of the solar system that focus primarily on its dynamical stability have considered the classical solar system, which assumes no additional forces or effects beyond Newtonian gravity, or they consider a constant strength of general relativity~\citep{Batygin2008,LaskarGastineau2009, Zeebe2015, Abbot2021, Abbot2022}.
We consider in more detail here how the dynamical stability of the solar system depends on the general relativistic perihelion precession rate of the planets. 

We do this by artificially adjusting the strength of the first order post-Newtonian correction over time, decreasing it from the values observed today to completely absent.
General modified versions of gravity have been developed to account for dynamical features seen in galactic and galactic cluster scales and are proposed to mimic the effect of dark matter \citep{Milgrom1983, MoffatToth2008, Verlinde2017}.
However, some of these theories result in changes to the perihelion precession rate of Mercury which are not compatible with current observations \citep{ChanLee2023}.
We investigate a decrease in the GR precession rate not because we think this mimics these modified theories of gravity, but because it offers a clean way to experiment with the dynamics of the solar system.

By comparing to previous work, we determine the effect the perihelion precession rate of Mercury from general relativistic corrections has on the probability that the system goes unstable.
The results of this work reconfirm the work of \cite{LaskarGastineau2009}, showing that the stability of the solar system is highly dependent on the presence of general relativity. 
What is new in this paper is that we also smoothly vary the strength of general relativistic corrections to the perihelion precession over the lifetime of the solar system and consider the implications for the very long-term evolution of the solar system. 
We develop a physical model based on a diffusion process \citep{Mogavero2021} that can explain the observed instability rate in our numerical experiments.
Our results show that a change in general relativistic corrections does not only change the planets' perihelion precession rate, but also the rate at which the precession rate diffuses with time.
We find that the dependence of the diffusion coefficient and the general relativistic corrections is smooth.
We find no evidence of any critical strength of GR that is required for stability. 

To do this, we begin with a discussion of Mercury's path to instability and the role of secular resonances in Section~\ref{sec:background}.
In Section~\ref{sec:numerics} we describe the numerical setup for our long-term ensemble of integrations of the solar system, how we modify the first order post-Newtonian corrections that effect the perihelion precession from general relativity, and the results of the integrations.
The diffusion model we use to interpret our results is developed and compared to previous work in Section~\ref{sec:model}.
In Section~\ref{sec:results} we compare the model to our new $N$-body data.
Finally, we close in Section~\ref{sec:conclusion} with the conclusions and implications.

\section{The path to instability}
\label{sec:background}
The secular evolution of the solar system was first described by Laplace and Lagrange \citep{Laplace1775, Laplace1776, Lagrange1776, Lagrange1778, Lagrange1781, Lagrange1782, Lagrange1783a, Lagrange1783b, Lagrange1784}.
See \cite{Laskar2013} for a comprehensive historical review.
This description of the solar system averages over the mean motions (phases) of the planets.
The ground-breaking linear expansion is a perturbation theory first order in the masses with a disturbing function second order in the eccentricities and inclinations. 
It exhibits no changes to the semi-major axis of the planets and over time shows only small changes to the eccentricities and inclinations, but not by an amount to allow for orbit crossing scenarios.
The solutions contain the perihelion precession of planetary orbits as well as the precession of the ascending nodes of the planets, but fundamentally they are perfectly periodic.
This initially seemed to \emph{prove} that the solar system is dynamically stable for all time, but these successive approximations did not provide rigorous bounds to actually prove it analytically \citep{Poincare1898}.

From this Laplace-Lagrange linear expansion, the fundamental eigenmodes for the variations in the eccentricities and inclinations of the solar system can be calculated \citep{Murray1999}.
When \cite{LeVerrier1840, LeVerrier1841} continued the perturbative expansion to higher order, significant terms emerged which make crucial corrections to the linear equations and showed that the linear theory could not be used for an indefinite period of time.
In spite of \cite{Poincare1899} proving the impossibility of an analytical solution to multi-planetary dynamics over an infinite time interval, quasi-periodic approximations were still the most accurate models at the time.
Further mathematical developments by \cite{Kolmogorov1954}, \cite{Arnold1963}, and \cite{Moser1962} (KAM) showed that for systems with a few degrees of freedom and in small regions around the initial conditions, the trajectories remain constrained to quasi-periodic solutions.
These KAM toridal regions are only isolating for systems with 2 degrees of freedom.
As such, for systems with more degrees of freedom, these constrained regions overlap and allow trajectories to pass into chaotic regions and effectively diffuse through phase space \citep{Laskar2013}.
With the aid of computer algebra systems, the integration of high-order secular expansions by \cite{Laskar1985, Laskar1986, Laskar1990} showed that the inner solar system is indeed chaotic.
Recent developments in high-order expansions continue to reveal insights into the dynamical richness of the solar system unachievable by $N$-body integrations alone \citep{Mogavero2021, Mogavero2022, HoangMogaveroLaskar2022}.

Over short timescales of $\sim 1\,\rm{Myr}$ the evolution of the secular frequencies is regular and can be calculated accurately either with computer aided expansion to higher order, or with direct $N$-body simulations. 
However, on timescales greater than $\sim 50\,\rm{Myr}$, the changes in secular frequencies are chaotic \citep{Laskar1990, LithwickWu2011}.
This chaos in the solar system is largely driven by overlapping secular resonances \citep{Laskar1989}.
Various resonances driving chaos in the inner solar system are at play, for example the Earth-Mars secular resonance $2(\varpi_4 - \varpi_3) - (\Omega_4 - \Omega_3)$ and the Mercury-Venus-Jupiter secular resonance $(\varpi_1-\varpi_5) - (\Omega_1 - \Omega_2)$ \citep{Laskar1992, SussmanWisdom1992},
where $\varpi_1, ..., \varpi_8$ are the longitude of the perihelia and $\Omega_1, ..., \Omega_8$ are the longitude of the ascending nodes of the planets.
Sometimes these secular resonances are also expressed in terms of the eigenfrequencies $g_1,..., g_8, s_1,..., s_8$ of the solar system associated with the precession of the perihelia and ascending nodes.
We follow the traditional convention to associate $g_1$ with Mercury, $g_2$ with Venus, and so forth, with our discussion focusing mainly on Mercury and Jupiter, and therefore $g_1$ and $g_5$.
Thus, the Earth-Mars resonance mentioned above would be $2(g_4-g_3)-(s_4-s_3)$.

Instabilities from the $g_1-g_5$ resonance can be seen in long-term solar system integration of $N$-body models
\citep{LaskarGastineau2009, Zeebe2015, Abbot2021, Abbot2022, BrownRein2022}. 
As the solar system evolves, the planets push each other into or out of secular resonances through exchanges of angular momentum \citep{Laskar2000, Zakamska2004}.
Reduced models help us to understand the nature of this process, specifically how the $g_1-g_5$ resonance affects the eccentricity pumping of Mercury, and how Mercury falls into the resonance in the first place after a diffusive walk through phase space \citep{LithwickWu2011, Batygin2015, Mogavero2021}.
Although the frequencies of the secular eigenmodes of the solar system are fixed in the simplest model, they do in fact show small variations.
In particular, the secular frequencies corresponding to the inner solar system change as a result of mutual interactions of the inner planets and the interactions of the inner planets with the outer gas giants \citep{Laskar1990}.
As we will show in Section~\ref{sec:model}, a simple diffusion process is often a remarkably accurate model that can explain most of the results seen in ensembles of direct $N$-body integrations. 

\section{\emph{N}-body simulations}
\label{sec:numerics}

\subsection{Methods}
\label{sec:setup}
We run direct $N$-body simulations of the solar system which we will later compare to our advection-diffusion model.
All of our simulations for the solar system (the Sun and eight planets) use exactly the same initial conditions which are taken from NASA JPL Horizons data at the J2000 epoch.
Even though all the initial conditions are the same, because the systems are chaotic and we use slightly different general relativistic corrections in each simulation, the simulations diverge quickly.
This is effectively the same as varying the initial conditions of one planet by a tiny amount \citep{LaskarGastineau2009}.
We integrate simulations forward in time using \reb \citep{ReinLiu2012} and the Wisdom-Holman integrator \whfast \citep{WisdomHolman1992,ReinTamayo2015} with symplectic correctors and the lazy implementation of the kernel method, \whckl \citep{Wisdom1996, Rein2019b}.
The \whckl integrator is well suited to provide highly accurate results for secularly evolving systems \citep{ReinTamayoBrown2019}.
We used a fixed timestep of $dt = \sqrt{11}\,\mathrm{days} \approx 3.317\,\mathrm{days}$.

The aim of our simulations is not to exactly reproduce the solar system, but to have a well defined model that can reproduce the most important dynamical effects.
Specifically, we do not consider the stellar evolution of the Sun even though we integrate beyond the end of its life on the main sequence.
We also neglect the effects resulting from the solar oblateness, moons, asteroids, tides, and other non-gravitational effects.

\subsection{Perihelion precession from general relativity}
\label{sec:precession}

Of all the contributions to Mercury's perihelion precession, the most important are the gravitational interactions from the other solar bodies contributing $g_{1,\rm planets} = 5.323\,''\mathrm{yr}^{-1}$ (arcseconds per year), followed by the precession from GR (the gravitoelectric effect) with $g_{1,\rm gr} = 0.4298\,''\mathrm{yr}^{-1}$, followed by other less significant contributors such as solar oblateness which provide $g_{1,\rm{other}} = 2.8\times10^{-4}\,''\mathrm{yr}^{-1}$ \citep{Park2017}.
Thus, the impact of general relativity follows closest behind Newtonian planet-planet interactions and is more than three orders of magnitude more influential than any additional effects.
As discussed in Section~\ref{sec:background}, the long-term stability of solar system is connected to Mercury's perihelion precession rate and thus the general relativistic effects are very important \citep{LaskarGastineau2009}.

In this paper, we consider an $N$-body Newtonian model of the solar system.
To have a perihelion precession that is consistent with general relativity, we include an additional potential term in our force calculation:
\begin{equation}
    \Phi_{\rm gr} = \alpha(t, \tau) \frac{6G^2M^2}{c^2r^2} \label{eq:grpot}
\end{equation}
where $r$ is the distance of the planet to the Sun and $M$, $G$, $c$ are the mass of the Sun, the gravitational constant, and the speed of light respectively. 
The perihelion precession rate caused by $\Phi_{\rm gr}$ is 
\begin{equation}
\label{eq:pomega-dot}
    \dot{\varpi}_{{}_{\mathrm{GR}}} = \alpha(t, \tau) \frac{3 G M n}{a c^2 (1-e^2)} 
\end{equation}
where $n = \sqrt{GM/a^3}$ is the mean motion of a planet with semi-major axis $a$ and $e$ is the planet's eccentricity.

The time dependent parameter $\alpha(t, \tau)$ in the above equations enables us to experiment with different strengths for the general relativistic corrections.
With $\alpha=1$, we refer to this potential as the standard first order post-Newtonian corrections.
This assumes that the Sun is the only body in solar system massive enough so that particles orbiting it experience general relativistic precession.
This commonly used model gets the general relativistic precession frequency right, but at the expense of introducing an error on the mean motions on the order of $\mathcal{O}(GM/ac^2)$ \citep{Nobili1986, SahaTremaine1994}.
For the discussion in this paper, the accuracy in the mean motion does not affect the results because of the relative importance of Mercury's perihelion precession rate to its stability compared to its mean motion (there are no mean motion resonances in the inner solar system).

With $\alpha=0$, we recover the purely classical regime with no general relativistic precession. 
Although this is clearly inconsistent with many observations that confirm general relativity, it can act as a useful case to study if we want to understand the dynamics of the solar system.
In previous work, \cite{LaskarGastineau2009} integrated thousands of solar system simulations for billions of years and showed that without general relativistic corrections, 60 per cent of the solutions resulted in an unstable solar system after 5 Gyr (defined as Mercury having an eccentricity beyond 0.9).
Conversely, when general relativistic corrections were included, only 1 per cent of the solutions were unstable \citep{LaskarGastineau2009}.
See also \fig{fig:laskar} below.

In this paper, we go beyond simply turning general relativistic precession on or off and allow for a time dependent parameter $\alpha$ defined as
\begin{equation}
\label{eq:alpha}
\alpha(t,\tau) = 
	\begin{cases}
	1 & t < 0\\
	1 - (t/\tau) & 0 \leq t \leq \tau\\
	0 & t > \tau
	\end{cases}\,.
\end{equation}
At the beginning of a simulation, $t=0$, the full GR corrections are taken into account, consistent with present day observations. After $t=\tau$ no more general relativistic corrections are applied and the simulation evolves classically.
We include the general relativistic corrections using \rebx \citep{Tamayo2020} and the \texttt{gr\_potential} module, modified to allow for the time varying form given in \eq{eq:grpot}.

\subsection{\emph{N}-body results}
\label{sec:stability}

\begin{figure}
    \centering
    \resizebox{0.99\columnwidth}{!}{\includegraphics{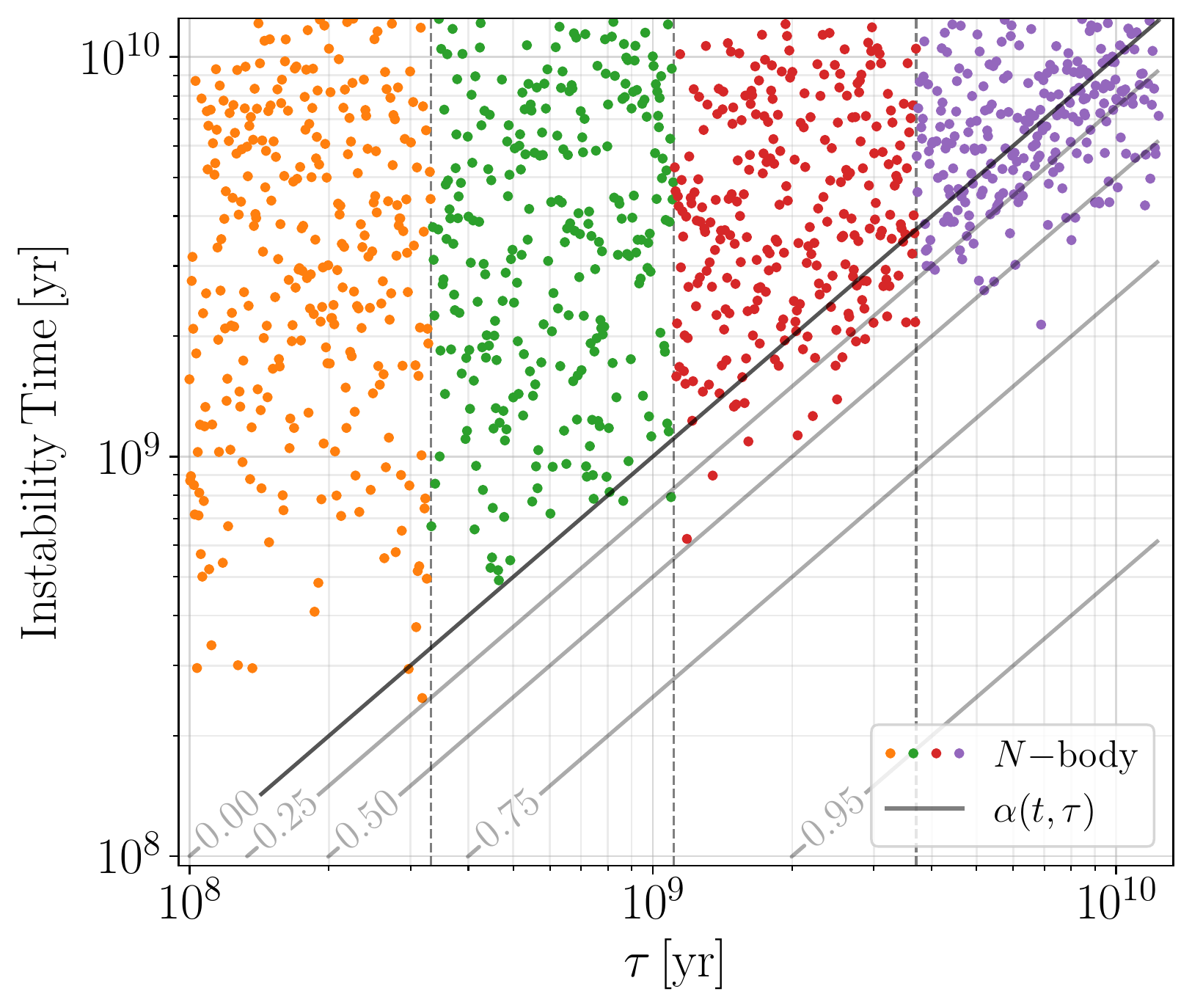}}
    \caption{The instability time of simulations in our ensemble that went unstable as a function of the control parameter $\tau$ which determines the time when general relativistic corrections to the perihelion precession are turned off.
    Simulations are significantly more likely to go unstable before 12.5 Gyr when no general relativistic corrections are present (upper left portion).
    The four different colours show the binning made for comparison to the model in Section~\ref{sec:results} and \fig{fig:survival_splits}.
    \label{fig:instability-time}
    }
\end{figure}

In this section we present the stability results of an ensemble of 1280 long-term integrations of the solar system.
We decrease the coupling strength, $\alpha(t,\tau)$, of the general relativistic corrections to perihelion precession linearly from the current value, 1, to 0 as defined in \eq{eq:alpha}.
For each simulation we assign a different value for $\tau$.
We sample $\tau$ log-uniformly between $0.1$ Gyrs and $12.5$ Gyrs.
We then integrate the simulations for $12.5$ Gyrs, or until a collision or escape event occurs.
We do not carry out any simulations beyond the first physical collision or escape.

We find that $1072$ out of $1280$ or 83.8 per cent of the simulations end up in an instability, all involving a Mercury-Venus close encounter or collision.
175 simulations go unstable while some fractional GR precession is still present.
We present the instability times with respect to $\tau$ in \fig{fig:instability-time}.
The instability time is shown on the $y$-axis, $\tau$ is shown on the $x$-axis.
The upper-most diagonal line is when $\alpha(t,\tau) = 0$ at $t = \tau$.
The remaining diagonal lines indicate the times when $\alpha(t,\tau)$ reaches $0.25, 0.5, 0.75,$ and $0.95$ (from top to bottom).

\section{Advection-diffusion model}
\label{sec:model}

\begin{figure}
    \centering
    \resizebox{0.99\columnwidth}{!}{\includegraphics[trim=0.1cm 0.1cm 0.1cm 0.1cm]{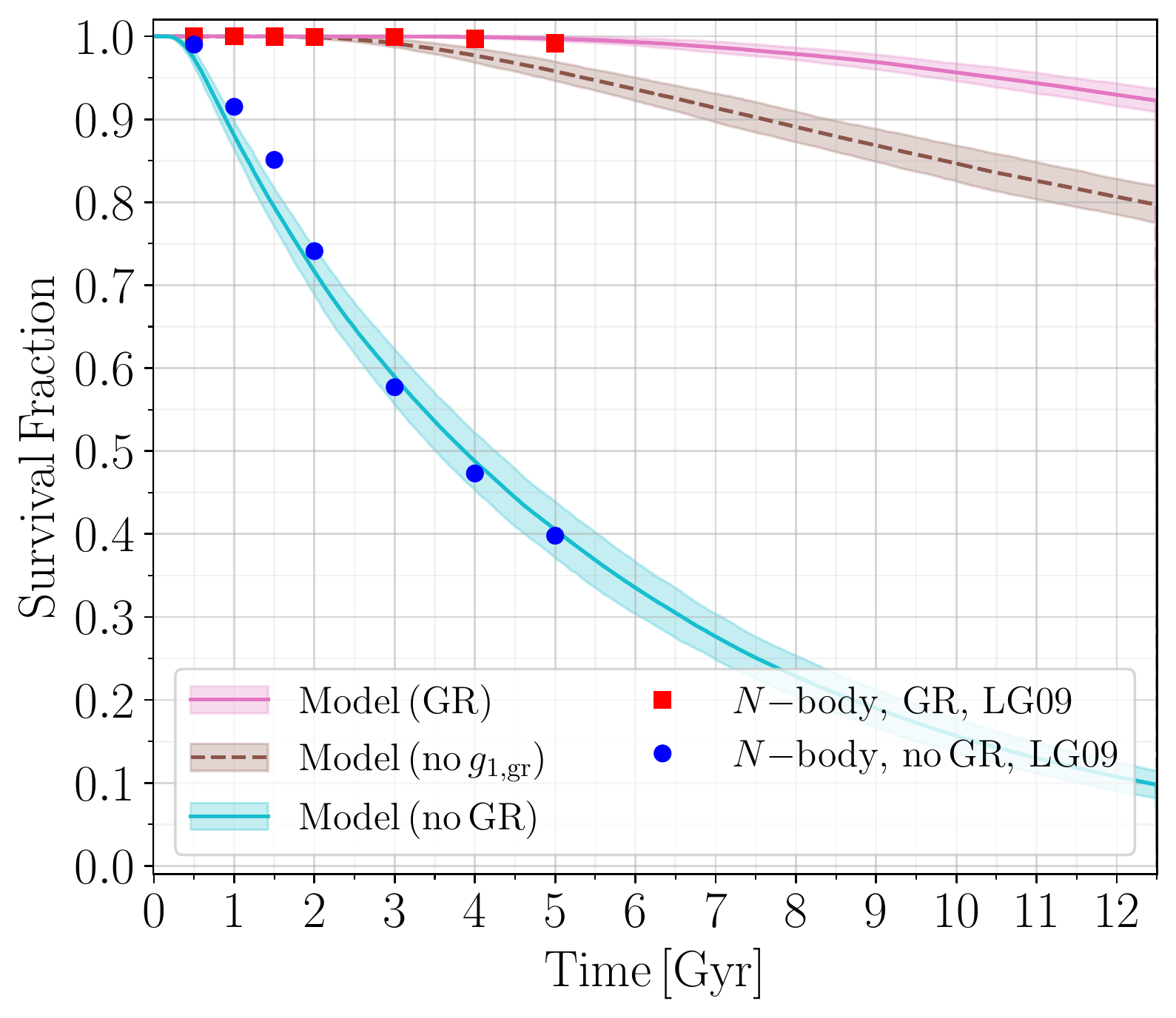}}
    \caption{A comparison showing the implementation of our model against the previous work of \protect\cite{LaskarGastineau2009}.
    The mean and $2\sigma$ uncertainties of the model is in good agreement with these numerical results in both cases with and without GR.
    The dashed curve shows the effect of changing the initial starting location of $g_1$ by removing $g_{1,\rm gr}$ from the GR model without altering the diffusion coefficient.
    The fact that this curve does not match the numerical experiments without GR reveals that the initial proximity between $g_1$ and $g_5$ alone does not account for the increase in instabilities.
    \label{fig:laskar}
    }
\end{figure}

\subsection{Fokker-Planck equation}

In this section, we develop a simple model that is physically motivated and can explain results from our new $N$-body simulations as well as the results from previous studies that have both included and excluded general relativistic corrections. 

Our model describes the evolution of Mercury's precession frequency $g_1$ as an advection-diffusion process. 
This can be seen as a natural extension of model used by \cite{Mogavero2021} who use a stochastic Wiener process to describe the evolution of Mercury's precession frequency $g_1$ (see their section 8.2).

We use the Fokker-Planck equation to describe the evolution of $p(g_1, t, \tau)$, the probability density of the secular frequency $g_1$ as a function of time $t$ and our model parameter $\tau$:
\begin{equation}
\label{eq:fokkerplanck}
    \frac{\partial p(g_1, t, \tau)}{\partial t} = -\mu(t,\tau)  \frac{\partial p(g_1, t, \tau)}{\partial g_1} + D(t, \tau)  \frac{\partial^2 p(g_1, t, \tau)}{\partial g_1^2}.
\end{equation}
Here $\mu(t,\tau)$ describes the advection and $D(t,\tau)$ the diffusion of $g_1$.
We use the current precession frequency of Mercury, $g_{1,0}=5.577''\mathrm{yr}^{-1}$, as the initial condition:
\begin{equation}
    p(g_1, 0, \tau) = \delta(g_1-g_{1,0}).
\end{equation}
There is an upper boundary which acts as a reflecting barrier at $g_{1,\mathrm{max}} = 5.72367''\mathrm{yr}^{-1}$ \citep{Mogavero2021}.
We also have a lower boundary where $g_1$ overlaps with $g_5= 4.257''\mathrm{yr}^{-1}$, the precession frequency of Jupiter. 
When this lower boundary is hit, the system is assumed to go unstable on a very short timescale because of the $g_1 - g_5$ secular resonance.

\subsection{Model with general relativistic precession}
In the limit where general relativistic corrections are present during the entire simulation, \cite{Mogavero2021} show that the results of high-order secular simulations can be closely matched with a constant diffusion coefficient $D_{\rm gr}$ and a vanishing convection term $\mu=0$.

A more intuitive understanding of these quantities can be achieved by rewriting the diffusion coefficient $D$ and the initial precession frequency $g_{1,0}$ into a diffusion timescale $T_D$ and a dimensionless parameter $\beta$ that describes how close the initial condition for $g_1$ is to the lower boundary $g_5$ where we expect an instability to occur. 
Making use of $(g_{1,\mathrm{max}} - g_5)$, the distance between the two boundaries, we can define the characteristic diffusion timescale as
\begin{equation}
    \label{eq:diffcoef}
    T_{D} = \frac{(g_{1,\mathrm{max}} - g_5)^2}{4 D}.
\end{equation}
and the parameter\footnote{This parameter is equivalent to the parameter $\alpha$ in \cite{Mogavero2021}.} $\beta$ that describes our initial conditions as:
\begin{equation}
    \beta = (g_{1,0} - g_5)/(g_{1,\mathrm{max}} - g_5).     \label{eq:beta}
\end{equation}
Using $g_5= 4.257''\mathrm{yr}^{-1}$ we have $\beta=0.9$.
\cite{Mogavero2021} show that a value of $T_{D,\rm gr} = 27.6\,\mathrm{Gyr}$ reproduces the statistical results from their simulations (which include general relativistic corrections) remarkably well. 
The survival fraction as a function of time corresponding to this model is the uppermost line plotted in \fig{fig:laskar}.
$N$-body data (an ensemble of 2501 simulations) from the previous work by \cite{LaskarGastineau2009} is overlaid as red squares for comparison.
It shows how this advection-diffusion model captures the instability rates of Mercury in $N$-body solar system simulations for the first 5 Gyr very well.
Obtaining a meaningful number of instabilities beyond the remaining lifetime of the sun is costly, but we show below that this model can also be used to give accurate results well beyond 5 Gyr.

\subsection{Model without general relativistic precession}
We can repeat the analysis from the previous section for the case without general relativistic precession.
In our notation, this is the case where $\tau\rightarrow 0$ and $\alpha=0$.
The contribution from general relativity to the precession frequency of Mercury is $g_{1,\rm gr} = 0.4298''\mathrm{yr}^{-1}$.
Without general relativity, our initial condition is therefore shifted to 
\begin{equation}
    p(g_1, 0, 0) = \delta(g_1- (g_{1,0}- g_{1,\rm gr}) ),
\end{equation}
or equivalently $\beta = 0.607$. 
If we assume this is the only change to the system, then we end up with a survival fraction shown as a dashed line in \fig{fig:laskar}. 
Clearly, although the rate of instability is higher, this does not fit the numerical experiments by \cite{LaskarGastineau2009}. 

This leads to the conclusion that we not only need to change the initial conditions, but also the diffusion coefficient.
To find the diffusion coefficient we can make use of the $N$-body results by \cite{LaskarGastineau2009} who find that 60 per cent of all simulations have gone unstable after 5 Gyr, using an ensemble of 201 simulations. 
Choosing $T_{D, \rm nogr} = 3.22\,\mathrm{Gyr}$ results in a model that fits those results well, as can be seen in \fig{fig:laskar}. 
While there are some discrepancies between the model and $N$-body results, especially in the 1-2 Gyr regime, we attribute this discrepancy to the relatively small number of simulations used in the ensemble.

\subsection{Model with time dependent general relativistic precession}
We are now in a position to describe a model for our case where we have a time dependent $\alpha(t,\tau)$ given by \eq{eq:alpha}, leading to a time dependent general relativistic precession.
In this paper, $\tau$ is the model parameter that describes the timescale on which the general relativistic precession changes.

For our model, the initial conditions are the present day precession frequency $g_{1,0}$, i.e.:
\begin{equation}
    p(g_1, 0, \tau) = \delta(g_1-g_{1,0}).
\end{equation}
The diffusion coefficient is no longer a constant, but a function of time
\begin{equation}
D(t, \tau) = 
\begin{cases}
	\left(\sqrt{D_{\rm gr}} \cdot \left(1- \frac{t}{\tau} \right)  
 + 	\sqrt{D_{\rm nogr}} \cdot \frac{t}{\tau}  \right)^2 
 & t \leq \tau\\
	D_{\rm no gr} & t > \tau
	\end{cases}
\end{equation}
We chose this specific form for the diffusion coefficient because the interpolation between $D_{\rm gr}$ and $D_{\rm nogr}$ corresponds to a simple linear interpolation in the forcing strength\footnote{Note that $D \sim t_c\cdot F^2$ where $t_c$ is the characteristic forcing timescale and $F$ is the characteristic forcing strength in a random walk process, see e.g. \cite{ReinPapaloizou2009}.}. 

In addition, we also now include the advection term in the Fokker-Planck equation. 
Specifically:
\begin{equation}
\mu(t, \tau) = 
\begin{cases}
	g_{1,\rm gr}/\tau & t \leq \tau\\
	0 & t > \tau
	\end{cases}.
\end{equation}
If we ignore the diffusion, then this form of $\mu$ moves the precession frequency from the initial value of  $g_{1,0}$ to  $g_{1,0} -  g_{1,gr}$ by time $t=\tau$, i.e. when $\alpha=0$ and there is no more general relativistic precession.

Note that there are no free parameters in this extension of the model. 
All the parameters, most importantly $D_{\rm gr}$ and $D_{\rm nogr}$ are already fixed by matching the simulations of \cite{LaskarGastineau2009}.

\section{Results}
\label{sec:results}

\begin{figure}
    \centering
    \resizebox{0.99\columnwidth}{!}{\includegraphics{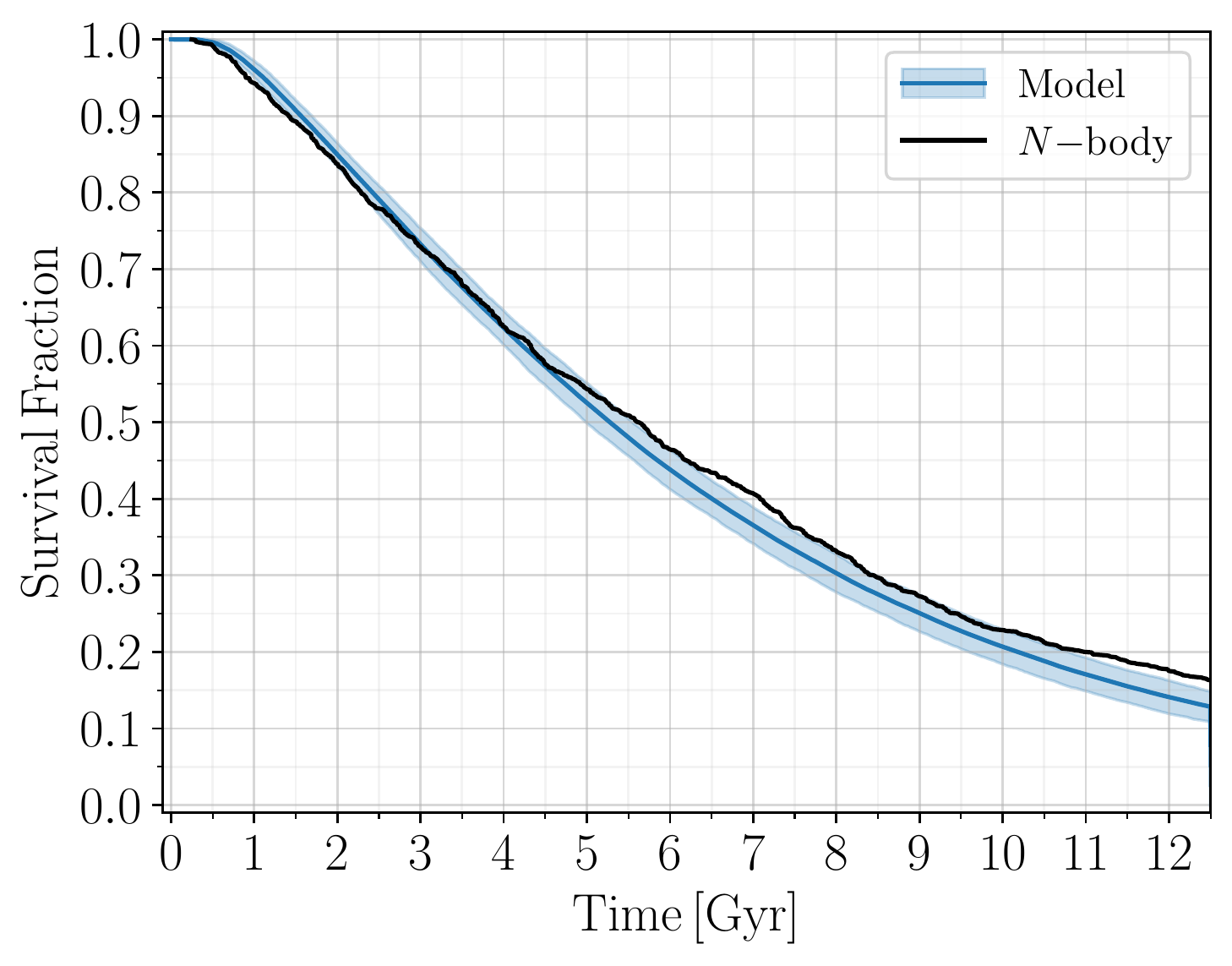}}
    \caption{The survival fraction of the entire ensemble of simulations compared to the survival given by the model described in Section~\ref{sec:model}.
    The mean and $2\sigma$ uncertainties are shown for the model.
    \label{fig:survival_all}
    }
\end{figure}

\begin{figure*}
    \centering
    \resizebox{0.99\textwidth}{!}{\includegraphics{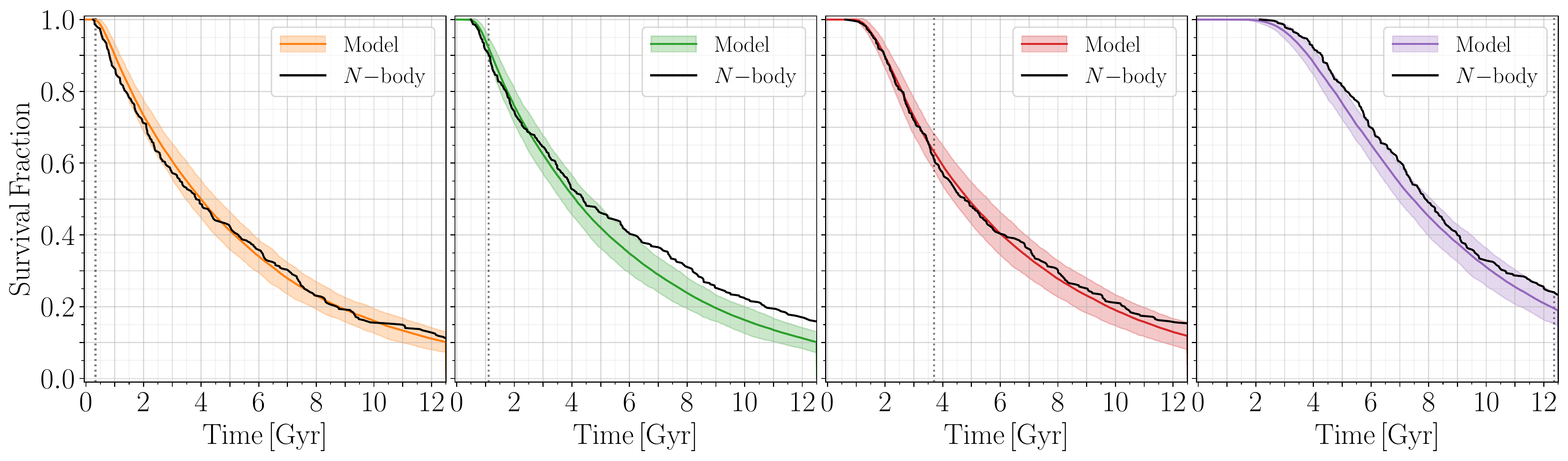}}
    \caption{The survival fraction of the ensemble of $N$-body simulations separated into four bins of $\tau$ as shown in \fig{fig:instability-time} is given in black.
    The mean survival of the model  described in Section~\ref{sec:model} is shown for each bin with $2\sigma$ uncertainties.
    The dotted vertical lines indicate when all of the $N$-body simulations in the bin have completely decoupled from general relativistic corrections.
    Thus, from $t=0$ to the dotted line, GR corrections are gradually being turned off.
    \label{fig:survival_splits}
    }
\end{figure*}

The survival fraction for our entire ensemble compared to the model is presented in \fig{fig:survival_all}.
It shows the results from our ensemble of 1280 $N$-body simulations compared against the advection-diffusion model with time dependent coefficients.
The mean and $2\sigma$ uncertainties expected for an ensemble size of 1280 are shown for the model.

For \fig{fig:survival_splits} we split our ensemble of simulations into four equally sized groups to show the dependence on $\tau$.
The colours match those used in \fig{fig:instability-time}.
Each panel includes the observed numerical survival fraction in our $N$-body simulations along with the expected mean survival fraction given by our model and the corresponding $2\sigma$ confidence interval for an ensemble size of 320.
The plots also include a dotted vertical line indicating when the last simulation in the bin has reached $\alpha(t,\tau) = 0$.
A notable feature in some of the panels is that the model produces instabilities slightly earlier than in the $N$-body experiments (starting around 4.5 Gyr).
Even so, the model and the numerical simulations agree remarkably well.
In particular, note that our model is not fit to the data, but rather uses a physically motivated model and parameters which are solely determined by the limiting cases done in previous work by \cite{LaskarGastineau2009} and \cite{Mogavero2021}.

The far left panel in \fig{fig:survival_splits} shows the survival fraction of simulations after removing the GR precession quickly\footnote{Quickly here means that the simulations in this bin all have the general relativistic corrections removed on a time scale shorter than the lower bound on the destabilization time of Mercury given by \cite{Mogavero2021} $\sim 0.56$~Gyr.}.
The results match those by \cite{LaskarGastineau2009}: $\sim 60$ per cent go unstable within 5~Gyr.
Note that removing GR precession from these integrations (even in the far future) does not immediately cause instability.
In fact, some simulations continue to remain stable (without GR) for more than $12$~Gyr.


\section{Conclusions}
\label{sec:conclusion}
In this paper, we explored how sensitive the stability of the solar system is to changes in the perihelion precession caused by general relativistic corrections.
Earlier work by \cite{LaskarGastineau2009} has shown that turning off general relativistic corrections can increase the fraction of instability within 5 Gyr by a factor of 60.
We develop a simple advection-diffusion model that can explain the results of previous $N$-body experiments which either include or exclude general relativistic precession, thus extending the model of \cite{Mogavero2021}. 
We then go beyond simply turning the GR precession on or off and instead vary the rate of GR precession smoothly over time.
We do this not because we think this is something that will occur in the real solar system, but because it offers a way to experiment with this dynamical system in a controlled way.

We show that our advection-diffusion model can naturally explain the case of a time varying general relativistic precession remarkably well without the introduction of any new free parameters. 
We are also confident that this model provides a sound extrapolation for the time to Mercury's dynamical instability beyond 5~Gyr (neglecting changes to the stellar life cycle of the Sun and other external factors).
This confidence is based on the work of others mentioned throughout this paper as well as the addition of our $N$-body ensemble extending all the way to 12.5 Gyr (to date the largest set of solar system integrations extended to this time).
The model shows strong agreement to $N$-body data for so many different regimes of general relativistic precession.
Additionally, the model reproduces the statistical results across different analytical and numerical methods, again, without the introduction of any new free parameters.

Without general relativistic corrections, the precession frequency of Mercury, $g_1$, is closer to that of Jupiter, $g_5$, by about $0.4298''\mathrm{yr}^{-1}$.
However, our results show that this alone does not explain the increased rate of instability observed in $N$-body experiments.
We show that our advection-diffusion model can represent the instability rate only if we significantly decrease the diffusion timescale (or increase the diffusion coefficient). 
In other words, although the $g_1 - g_5$ resonance determines when Mercury's eccentricity reaches a critical value, the process to get to this resonance is much faster without GR corrections. 

This result might not be surprising, given that the solar system is a complex chaotic system with way more degrees of freedom than our model (which just allows for a diffusion in $g_1$). 
Our simulations with time-varying general relativistic precession allow us to gain some more insight into this process. 
As we've shown in Section~\ref{sec:results}, our model matches the $N$-body experiments very well given there are no free parameters aside from those used to match the limiting cases of GR being completely on or off. 
This leads to several conclusions. 
First, there is no critical value of the general relativistic precession rate that needs to be crossed in order to significantly change the instability rate.
Second, because this is such a smooth transition all the way from one limit to the other, we can rule out a process that involves a sweeping secular resonance.
If this were not so smooth, we would see an increase in the instability rate at some critical GR precession frequency.
This makes sense because the diffusion in $g_1$ is strong enough so that a small change due to the GR precession has a negligent effect. 
This provides additional evidence that the stability of the solar system is robust to moderate changes to the secular system \citep[see also][]{Laskar1990, BrownRein2022}.
The fact that we are able to model the evolution so well with a simple diffusion model also shows that current numerical results \citep{LaskarGastineau2009, Zeebe2015, Abbot2021} are robust against small perturbations whether they are physical or numerical \citep{Abbot2022}.
We expect that statistical results are in agreement as long as simulations resolve secular frequencies accurately enough so that physical diffusion (not numerical diffusion or advection) is the dominant driver of the instability.

In summary, we reconfirm that the solar system's evolution is well described by a simple (advection-)diffusion process, even in the presence of other small perturbations.
Here we considered the perturbation to be the change in the general relativistic precession frequency, but any other physical effect that changes the planets' precession frequencies will lead to a similar result.
The initial motivation of this project was to use the stability of the solar system as a test-bed for additional physics that are typically not included in simulations of the solar system. 
For example, had our results shown that slowly changing the general relativistic precession frequency leads to significant increase in the instability rate, then we could have placed a limit on alternative theories of general relativity by noting that the solar system has not gone unstable yet. 
Clearly this is not the case. 
For future work, we propose further investigation into the expected time to instability for other secularly evolving planetary systems and whether or not a one dimensional advection-diffusion model is representative of more systems or only the solar system \citep[see e.g.][]{Hussain2020}.

\section*{Data availability}
The data underlying this article will be shared on reasonable request to the corresponding author.

\section*{Code availability}
A repository containing a portion of the data underlying this article and code for running the simulations and generating the figures can be found at \href{https://github.com/zyrxvo/GR-and-Long-term-Stability}{github.com/zyrxvo/GR-and-Long-term-Stability}.

\definecolor{lime}{HTML}{A6CE39}
\DeclareRobustCommand{\orcidicon}{%
	\begin{tikzpicture}
	\draw[lime, fill=lime] (0,0) 
	circle [radius=0.16] 
	node[white] {{\fontfamily{qag}\selectfont \tiny ID}};
	\draw[white, fill=white] (-0.0625,0.095) 
	circle [radius=0.007];
	\end{tikzpicture}
	\hspace{-2mm}
}
\foreach \x in {A, ..., Z}{%
	\expandafter\xdef\csname orcid\x\endcsname{\noexpand\href{https://orcid.org/\csname orcidauthor\x\endcsname}{\noexpand\orcidicon}}
}
\newcommand{\orcidauthorA}{0000-0002-9354-3551}
\newcommand{\orcidauthorB}{0000-0003-1927-731X}

\section*{ORCID iDs}
Garett Brown \orcidA{} \href{https://orcid.org/0000-0002-9354-3551}{https://orcid.org/0000-0002-9354-3551}
Hanno Rein \orcidB{} \href{https://orcid.org/0000-0003-1927-731X}{https://orcid.org/0000-0003-1927-731X}
 
\section*{Acknowledgments}
We are very grateful to an anonymous referee for a helpful review that improved the quality of this paper.
We would like to thank Scott Tremaine, Sam Hadden, Dang Pham, and Mykhaylo Plotnykov for useful discussions. 
This research has been supported by the NSERC Discovery Grants RGPIN-2014-04553 and RGPIN-2020-04513.
This research was made possible by the open-source projects 
\texttt{Jupyter} \citep{jupyter}, \texttt{iPython} \citep{ipython}, \texttt{matplotlib} \citep{matplotlib, matplotlib2}, and \texttt{GNU Parallel} \citep{gnuparallel}.
This research was enabled in part by support provided by Digital Research Alliance of Canada (formerly Compute Canada) (\href{https://alliancecan.ca/en}{alliancecan.ca}).
Computations were performed on the Niagara supercomputer \citep{SciNet2010, Ponce2019} at the SciNet HPC Consortium (\href{www.scinethpc.ca}{scinethpc.ca}). 
SciNet is funded by: the Canada Foundation for Innovation; the Government of Ontario; Ontario Research Fund - Research Excellence; and the University of Toronto.


\bibliography{full}

\end{document}